# An Isotropic to Anisotropic Transition in a Fractional Quantum Hall State

Michael Mulligan$^a$, Chetan Nayak$^{b,c}$, and Shamit Kachru$^{b,d,1}$

$^a$*Center for Theoretical Physics, MIT, Cambridge, MA 02139, USA*
$^b$*Department of Physics, University of California, Santa Barbara, CA 93106, USA*
$^c$*Microsoft Station Q, Santa Barbara, CA 93106, USA*
$^d$*Kavli Institute for Theoretical Physics, Santa Barbara, CA 93106, USA*

Email: mcmullig@mit.edu, nayak@kitp.ucsb.edu, skachru@kitp.ucsb.edu

We study a novel abelian gauge theory in 2+1 dimensions which has surprising theoretical and phenomenological features. The theory has a vanishing coefficient for the square of the electric field $e_i^2$, characteristic of a quantum critical point with dynamical critical exponent $z = 2$, and a level-$k$ Chern-Simons coupling, which is *marginal* at this critical point. For $k = 0$, this theory is dual to a free $z = 2$ scalar field theory describing a quantum Lifshitz transition, but $k \neq 0$ renders the scalar description non-local. The $k \neq 0$ theory exhibits properties intermediate between the (topological) pure Chern-Simons theory and the scalar theory. For instance, the Chern-Simons term does not make the gauge field massive. Nevertheless, there are chiral edge modes when the theory is placed on a space with boundary, and a non-trivial ground state degeneracy $k^g$ when it is placed on a finite-size Riemann surface of genus $g$. The coefficient of $e_i^2$ is the only relevant coupling; it tunes the system through a quantum phase transition between an isotropic fractional quantum Hall state and an anisotropic fractional quantum Hall state. We compute zero-temperature transport coefficients in both phases and at the critical point, and comment briefly on the relevance of our results to recent experiments.

April, 2010

---

[1] On leave of absence from Department of Physics, Stanford University and SLAC.

## 1. Introduction

In 2+1 dimensions, standard arguments of effective field theory suggest that in the presence of parity (P) and time-reversal invariance (T) violations, the infrared dynamics of a gauge field should be controlled by the Chern-Simons Lagrangian. Even in the abelian case, this action contains a great deal of physics.

In the context of the fractional quantum Hall effect (FQHE) on the plane, where the background magnetic field breaks P and T, the coefficient of the Chern-Simons action encodes the Hall conductance, and governs the charge and statistics of the quasiparticle excitations (see e.g. [1,2]). Furthermore, when the system is placed on a higher genus Riemann surface or a surface with boundaries, the existence of degenerate ground states (split by exponentially small corrections in the system size) and gapless boundary excitations follows directly [3].

The argument that the Chern-Simons term is the single most relevant term involving only a gauge field implicitly assumes the conventional scaling of the coordinates and fields (that occurs in e.g. relativistic systems). More generally, one can imagine scale-invariant theories with a dynamical critical exponent $z$. In such a theory, under a scale transformation controlled by a parameter $\lambda$,

$$t \to \lambda^z t, \quad x \to \lambda x \tag{1.1}$$

with the components of the gauge field scaling as above but with inverse powers of $\lambda$. Choosing $z = 1$, which is also consistent with Lorentz invariance, the Chern-Simons term is marginal and the leading irrelevant operator is the Maxwell term for the gauge field. The theory flows to pure Chern-Simons theory in the deep infrared (IR). However, the Chern-Simons term itself is topological, and does not give a preferred choice of scaling; it is marginal for any choice of $z$. It is therefore natural to ask how this picture is modified when general $z$ is allowed – especially in the context of condensed matter systems, where Lorentz invariance is not a fundamental symmetry. Here, we consider abelian Chern-Simons theory and focus on the specific case $z = 2$.

*The $z = 2$ abelian gauge theory of the quantum Lifshitz transition*

It is conventional to begin the study of electron systems, e.g. the Hall system, by assuming the existence of a conserved (electromagnetic) current. Thus, we study a system described by a conserved current $J$. Conservation of the current allows one to introduce a gauge field, $a_\mu$,

$$J_\mu = \frac{1}{2\pi}\epsilon_{\mu\nu\rho}\partial_\nu a_\rho. \tag{1.2}$$

We use the lower case $a_\mu$ to distinguish this gauge field from the electromagnetic field $A_\mu$. We will later couple $J_\mu$ to the electromagnetic field $A_\mu$ in order to compute response functions. Greek indices run over all three coordinates, $t$ signifies the temporal direction, and Roman indices stand for the two spatial directions. Since the flux of this gauge field is the electric charge, $J_0 = \frac{1}{2\pi}\epsilon_{ij}\partial_i a_j$, the gauge field must be non-compact. Otherwise, instantons could change the flux of $a_i$ by multiples of $2\pi$, which would violate conservation of electric charge.

The $z = 2$ effective action we consider has a Chern-Simons term, as we discuss below. When the Chern-Simons term has vanishing coefficient, the action is:

$$S = \frac{1}{g^2}\int dt\, d^2x \left[e_i \partial_t a_i + a_t \partial_i e_i - H[e,a]\right], \tag{1.3}$$

where the Hamiltonian $H[e,a]$ equals

$$H[e,a] = \frac{\kappa^2}{2}(\partial_i e_j)^2 + \frac{1}{2}b^2, \tag{1.4}$$

with a 'magnetic field' b given by

$$b = \epsilon_{ij}\partial_i a_j. \tag{1.5}$$

The reason for the quotation marks is that $J_\mu = \frac{1}{2\pi}\epsilon_{\mu\nu\rho}\partial_\nu a_\rho$ implies that $b$ is actually the charge density. One might have expected $e_i\partial_t a_i + a_t\partial_i e_i$ to have a different coefficient than $b^2/2$, but we can rescale $e_i$ to make the two coefficients the same. The action (1.3) is written in first-order phase space formalism where the electric field $e_i$ and vector field $a_\mu$ are treated as independent variables. (1.3) is invariant under the following $z = 2$ scaling of coordinates and fields:

$$t \to \lambda^2 t, \quad x \to \lambda x, \quad a_t \to \lambda^{-2} a_t, \quad a_i \to \lambda^{-1} a_i, \quad e_i \to \lambda^{-1} e_i. \tag{1.6}$$

Because the action is quadratic, it is possible to integrate out the electric field and write the action solely in terms of the gauge field. This gauge field has quadratic dispersion. (1.3) is called the quantum Lifshitz action and describes a $z = 2$ gauge field.

The $e_i^2$ operator is relevant – dominating in the IR – and its coefficient must be fine-tuned to zero in order to sit at the critical point described by (1.3). If its coefficient (in the Hamiltonian) is positive, the $(\partial_i e_j)^2$ operator can be ignored. One can then integrate out the electric field, thereby recovering a Maxwell theory with the 'speed of light' determined by the magnitude of the coefficient of the $e_i^2$ term. If the coefficient of $e_i^2$ is negative (in which case, a higher-order term such as $(e_i^2)^2$ must be present with positive coefficient to stabilize the theory), the system is in an anisotropic phase with spontaneously broken rotational symmetry. We discuss a few aspects of this phase later in the more general setting in which the Chern-Simons term has arbitrary coefficient.

The theory also admits the marginal operator $(e_i^2)^2$. This operator violates the shift symmetry $e \to e + \text{const}$. (For this reason, it does not occur in the effective theory of the square lattice quantum dimer model [4] at the Rokhsar-Kivelson point [5].) As noted in the previous paragraph, such a term can stabilize the theory if the coefficient of $e_i^2$ is negative. If $(e_i^2)^2$ is added to the Hamiltonian with a negative coefficient, the Hamiltonian is unbounded from below and the theory has no vacuum unless a higher-order term such as $(e^2)^3$ term is present with positive coefficient. If $(e_i^2)^2$ is present with a positive coefficient, it is known to be marginally *irrelevant* from computations in [4,6].

In 2+1 dimensions, a gauge field may be dualized into a scalar field. In the context of $z = 2$ scaling, we introduce a scalar field by writing $J_\mu = (\partial_t \phi, \partial_i \nabla^2 \phi)$. The conservation equation for $J_\mu$ follows from the $\phi$ field equation derived from the Lagrangian,

$$L = \frac{1}{2} \int dt d^2 x \Big[ (\partial_t \phi)^2 - \kappa^2 (\nabla^2 \phi)^2 \Big]. \tag{1.7}$$

$\phi$ has scaling dimension zero and we have introduced the marginal parameter $\kappa^2$. In this language the part of the relevant $e^2$ deformation is played by the $(\nabla \phi)^2$ operator.

The scalar form of the action (1.7) describes a Lifshitz scalar. (1.7) has been studied in the context of phase transitions in both magnetically [7] and topologically ordered [8] systems. It has also inspired a generalization of gauge/gravity duality to systems with dynamical scale invariance [9]. Further, it provides a simple 2+1 dimensional example of a theory that displays universal subleading terms in the expression for the entanglement entropy [10]. See [11] for a calculation of universal subleading terms at the free-field and Wilson-Fisher fixed points of the $O(N)$ model. It would be interesting to consider the behavior of the entanglement entropy as a function of $e^2$ coefficient in the theory considered in this paper.

*Adding the Chern-Simons term*

The action (1.3) preserves parity and time-reversal invariance. The Chern-Simons operator, which explicitly breaks these symmetries, may be appended to the action:

$$S = \int dt\, d^2x \left[ \frac{1}{g^2} \left( e_i \partial_t a_i + a_t \partial_i e_i - H[e,a] \right) + \frac{k}{4\pi} \epsilon_{\mu\nu\lambda} a_\mu \partial_\nu a_\lambda \right]. \quad (1.8)$$

The Chern-Simons term is *marginal* under the scaling (1.6), and therefore competes on equal footing with the Lifshitz terms in determining the IR dynamics. This is the first surprise, and it should be contrasted with the situation in Maxwell-Chern-Simons theory, where the Chern-Simons term renders the gauge field massive. To get some intuition into the effect of the Chern-Simons term on the Lifshitz theory, it is useful to compute the propagators of this quadratic theory:

$$\begin{aligned}
\langle e_i(-i\omega_n, -\mathbf{p})\, e_j(i\omega_n, \mathbf{p}) \rangle &= -\frac{g^2 p^2}{\omega_n^2 + \tilde{\kappa}^2 p^4} P^T_{ij}(p) - \frac{g^6 \kappa^2 \tilde{k}^2 p^2 \delta_{ij} + g^4 \tilde{k} \omega_n \epsilon_{ij}}{\omega_n^2 + \tilde{\kappa}^2 p^4} \\
\langle e_i(-i\omega_n, -\mathbf{p})\, a_j(i\omega_n, \mathbf{p}) \rangle &= -\frac{g^2 \omega_n}{\omega_n^2 + \tilde{\kappa}^2 p^4} P^T_{ij}(p) - \frac{g^4 \kappa^2 \tilde{k} p_i \epsilon_{jk} p_k}{\omega_n^2 + \tilde{\kappa}^2 p^4} \\
\langle e_i(-i\omega_n, -\mathbf{p})\, a_t(i\omega_n, \mathbf{p}) \rangle &= \frac{-ig^2 \left( (\omega_n^2 + \kappa^2 p^4) p_i - \kappa^2 g^2 \tilde{k} \omega_n \epsilon_{ij} p_j \right)}{p^2 (\omega_n^2 + \tilde{\kappa} p^4)} \\
\langle a_i(-i\omega_n, -\mathbf{p})\, a_j(i\omega_n, \mathbf{p}) \rangle &= -\frac{g^2 \kappa^2 p^2}{\omega_n^2 + \tilde{\kappa}^2 p^4} P^T_{ij}(p) - \frac{g^2}{p^2} \left( \frac{p_i p_j}{p^2} \right) \\
\langle a_t(-i\omega_n, -\mathbf{p})\, a_i(i\omega_n, \mathbf{p}) \rangle &= \frac{ig^2 \omega_n p_i}{p^4} - \frac{i\kappa^4 g^4 \tilde{k} \epsilon_{ij} p_j p^6}{p^4 (\omega_n^2 + \tilde{\kappa}^2 p^4)} \\
\langle a_t(-i\omega_n, -\mathbf{p})\, a_t(i\omega_n, \mathbf{p}) \rangle &= \frac{g^2 (\omega_n^4 + (\kappa^2 + \tilde{\kappa}^2) \omega_n^2 p^4 + \kappa^4 p^8)}{p^4 (\omega_n^2 + \tilde{\kappa}^2 p^4)}
\end{aligned} \quad (1.9)$$

where $P^T_{ij}(p) = \delta_{ij} - p_i p_j / p^2$ is the transverse projector, $\tilde{k} \equiv k/2\pi$, and $\tilde{\kappa}^2 = \kappa^2 (1 + \kappa^2 g^4 \tilde{k}^2)$. In (1.9), we have fixed the gauge by inserting a delta function $\delta(\partial_i a_i - \omega)$ into the action and integrating $\int d\omega\, \exp(-\frac{1}{2\xi} \int \omega^2)$. For simplicity, we have taken $\xi = 1$; radiation gauge $\partial_i a_i = 0$ is $\lambda = 0$. The Faddeev-Popov ghosts decouple and have been dropped. As may be seen from (1.9), the Chern-Simons term modifies the propagators from their form in the pure Lifshitz theory, but does not make them massive.

Besides the $e_i^2$ term, there are no other relevant or marginal operators consistent with gauge invariance, spatial $SO(2)$ rotation invariance, and temporal and spatial translation symmetry. This is true even when parity and time-reversal invariance are broken. One might worry about the fate of the marginally irrelevant $(e^2)^2$ operator mentioned above. If we add this term with coefficient $\lambda$ to the Lagrangian then the RG equation for $\lambda$ is

$$\frac{d\lambda}{d\ell} = -\frac{\lambda^2 g^4}{2\pi \tilde{\kappa}^3} \left( \frac{9}{2} + 16 g^4 \tilde{k}^2 \tilde{\kappa}^2 \right). \quad (1.10)$$

Thus, the operator is marginally irrelevant. If a system has only a discrete rotational symmetry dictated by the underlying lattice, then there will be additional operators which are marginal at tree-level (and run at one-loop for $k = 0$ [4,6]). However, we have in mind systems in which the lattice plays little role and underlying spatial anisotropies are very small. Thus, they can be ignored except, perhaps, at extremely low energies, where even a tiny initial anisotropy may flow to larger values.

Thus, in this paper, we study the gauge field action

$$S = \int dt\, d^2 x \left[ \frac{1}{g^2}\Big(e_i \partial_t a_i + a_t \partial_i e_i - \frac{r}{2}e_i^2 - \frac{\kappa^2}{2}(\partial_i e_j)^2 - \frac{\lambda}{4}(e_i^2)^2 - \frac{1}{2}b^2\Big) + \frac{k}{4\pi}\epsilon_{\mu\nu\rho} a_\mu \partial_\nu a_\rho \right]. \tag{1.11}$$

For $r > 0$, we can integrate out $e_i$, thereby obtaining Maxwell-Chern-Simons theory, which describes a quantum Hall state. As we will describe in Sec. 3, for $r < 0$, the effective action (1.11) describes an anisotropic state. The nature of this anisotropic state depends on whether or not the underlying system is invariant under $SO(2)$ rotational symmetry. If it is, then the state spontaneously breaks the $SO(2)$ symmetry. It is superconducting in one direction and insulating in the other. Otherwise, the state is a quantum Hall state with anisotropic finite-frequency longitudinal conductivities reflective of the intrinsic anisotropy of the system. Our main focus is the critical point at $r = 0$ between these two phases. Here and henceforth, we call this $z = 2$ gauge field action 'Lifshitz-Chern-Simons theory'. We will also analyze the $r < 0$ anisotropic phase in some detail.

We focus on the gauge theory language because there is no local expression for the Chern-Simons coupling in the scalar formulation of the theory.[2] There are two perspectives on this Lifshitz-Chern-Simons theory. From the viewpoint of the undeformed abelian Lifshitz theory (1.3) which preserves parity and time-reversal invariance, it is interesting to ask how the Chern-Simons 'deformation' affects the gapless $z = 2$ modes. Since the Chern-Simons coupling must be quantized (as we see below), it cannot be viewed as a small deformation, but it is nevertheless useful to view it as a deformation of the Lifshitz theory. Alternatively, we may view the action (1.3) as a deformation of a Maxwell-Chern-Simons theory. Again, this is not a small deformation but, rather, one which is large enough to drive the coefficient of the Maxwell term to zero (while keeping the coefficient of $(\partial_i e_j)^2$ positive), thereby driving the system to a critical point. Passing through this critical point, the system leaves the topological phase described in the IR by pure Chern-Simons theory and goes into an anisotropic phase.

---

[2] We will describe how to map the abelian Lifshitz-Chern-Simons gauge theory onto a scalar field theory by a $non - local$ change of variables in Sec. 2.

The structure of this paper is as follows. In Sec. 2.1, we canonically quantize the Lifshitz-Chern-Simons theory and find that the $z = 2$ gapless modes persist. However, these modes acquire a parity-violating unit of spin. In Sec. 2.2, we put the theory on a nontrivial Riemann surface and find that despite the presence of gapless bulk modes, the non-trivial ground-state degeneracy (as defined carefully in a finite-size system) of the model persists. In Sec. 2.3, we argue that when the theory is placed on a surface with boundary, we find gapless chiral edge modes which persist in the presence of disorder. In Sec. 3.1, we describe the anisotropic phase which arises when the $e_i^2$ term appears in the Hamiltonian with the "wrong sign" in a clean system. In Sec. 3.2, we describe this phase in the presence of disorder. Finally, in Sec. 4, we couple an external gauge field to the Lifshitz-Chern-Simons theory in order to calculate various response functions. We describe the phase diagram as a function of the single relevant $e^2$ perturbation.

## 2. Critical Point

### 2.1. Canonical Quantization of the Critical Theory

Before studying the Lifshitz-Chern-Simons theory, we recall the excitation spectrum for Maxwell-Chern-Simons theory on $R^3$. In [12], Deser, Jackiw, and Templeton find that the Chern-Simons coupling imparts a tree-level mass for the linearly dispersing gauge field excitation. This photon obtains a mass proportional to the Chern-Simons level multiplied by the ultraviolet (UV) cutoff and carries a unit of spin, reflecting the violation of parity. Below, we closely follow the quantization procedure discussed in [12].

The Lifshitz-Chern-Simons action is

$$S = \int dt\, d^2x \left[ \frac{1}{g^2} \left( e_i \partial_t a_i + a_t \partial_i e_i - \frac{\kappa^2}{2} (\partial_i e_j)^2 - \frac{1}{2} b^2 \right) + \frac{k}{4\pi} \epsilon_{\mu\nu\rho} a_\mu \partial_\nu a_\rho \right]. \tag{2.1}$$

The Chern-Simons level $k$ is quantized when one considers the theory on non-trivial three-manifolds [13]. Note that all terms in this action scale in the same way, in stark contrast with Maxwell-Chern-Simons theory, where the Chern-Simons term is more relevant than the Maxwell term and gives the gauge field a mass. Since the theory is critical, no UV cutoff is required to make the couplings dimensionless. (In contrast, a UV cutoff is required in studies of the Maxwell-Chern-Simons Lagrangian). However, the natural cutoff for this theory is the single-electron gap, above which the action (2.1) is clearly incomplete.

After formally integrating out the electric field (which is well-defined as long as the momenta are non-zero), we obtain the Lagrangian

$$S = \frac{1}{2} \int dt\, d^2x \left[ \frac{1}{g^2} \left( \kappa^2 (\partial_i e_j)^2 - b^2 \right) + \frac{k}{2\pi} \epsilon_{\mu\nu\rho} a_\mu \partial_\nu a_\rho \right], \tag{2.2}$$

where

$$e_i = \left(\frac{1}{\kappa^2}\right) \frac{\partial_i a_t - \partial_t a_i}{\nabla^2}, \quad b = \partial_1 a_2 - \partial_2 a_1. \tag{2.3}$$

In terms of the canonical momenta $\Pi_i$

$$\Pi_i = \frac{\delta S}{\delta(\partial_t a_i)} = \frac{e_i}{g^2} + \frac{k}{4\pi} \epsilon_{ij} a_j,$$

the Hamiltonian is

$$H = \frac{1}{2} \int d^2x \left[ \kappa^2 g^2 (\partial_i \Pi_j - \frac{k}{4\pi} \epsilon_{jk} \partial_i a_k)^2 + \frac{1}{g^2} b^2 \right]. \tag{2.4}$$

Canonical quantization of this theory is simplest in the gauge in which we take $a_t = 0$, restrict to transverse fluctuations of the gauge field $\partial_i a_i = 0$, and impose the canonical commutation relations

$$[a_i(x), \Pi_j(y)] = i\delta_{ij}\delta(x-y),$$

where we have chosen $\Pi_i = -i\delta/\delta a_i$. Gauss' law, which is obtained from the $a_t$ equation of motion

$$G = \partial_i \Pi_i + \frac{k}{4\pi} b = 0 ,$$

must be imposed as a constraint on physical states $G|\Psi\rangle = 0$. Note that $G$ commutes with the Hamiltonian. Thus, once imposed on a state, Gauss' law is satisfied by subsequent evolution.

A general Schrödinger wave functional $\Psi(a)$ has the form

$$\Psi(a) = e^{\frac{ik}{4\pi} \int d^2x\, b \frac{\partial_k a_k}{\partial^2}} \Phi[a_T^2] \tag{2.5}$$

for some functional $\Phi[a_T^2]$ of the transverse field $a_T$, because

$$\left( \partial_j \frac{-i\delta}{\delta a_j} + \frac{k}{4\pi} b \right) e^{\frac{ik}{4\pi} \int d^2x\, b \frac{\partial_k a_k}{\partial^2}} = 0$$

and
$$\partial_j \frac{-i\delta}{\delta a_j} \Phi[a_T^2] = -2i\Phi'[a_T^2]\partial_j a_T^j = 0.$$

The phase prefactor ensures Gauss' law is satisfied as long as $\Phi$ is a functional of the transverse gauge field. To find the spectrum of excitations of $\Phi$, it's convenient to conjugate the Hamiltonian by the phase prefactor,

$$\begin{aligned} H_T &= e^{\frac{-ik}{4\pi} \int d^2x\, b \frac{\partial_j a_j}{\partial^2}} H e^{\frac{ik}{4\pi} \int d^2x\, b \frac{\partial_k a_k}{\partial^2}} \\ &= \frac{1}{2} \int d^2x \left[ \kappa^2 g^2 (\partial_i \Pi_T^j)^2 + \left( \frac{1}{g^2} + \frac{\kappa^2 k^2 g^2}{4\pi} \right)(\partial_i a_T^j)^2 \right], \end{aligned} \qquad (2.6)$$

when acting on functionals of the transverse field $a_T$. Note that the partial derivatives in the Hamiltonian commute with the unitary transformation $H \to H_T$.

Hamilton's equations are

$$\begin{aligned} \partial_t a_T^i &= \frac{\delta H_T}{\delta \Pi_T^i} = -\kappa^2 g^2 \nabla^2 \Pi_T^i, \\ \partial_t \Pi_i &= -\frac{\delta H_T}{\delta a_T^i} = \left( \frac{1}{g^2} + \frac{\kappa^2 k^2 g^2}{4\pi} \right) \nabla^2 a_T^i. \end{aligned} \qquad (2.7)$$

Differentiating the first equation with respect to $t$ and plugging the result into $\nabla^2$ of the second we find

$$\partial_t^2 a_T^i + \kappa^2 \left(1 + \frac{\kappa^2 k^2 g^4}{4\pi}\right)(\nabla^2)^2 a_T^i = 0 , \qquad (2.8)$$

which is precisely the equation obeyed by a Lifshitz scalar field with quadratic dispersion (1.7). The constraint $\partial_i a_T^i = 0$ relates the two spatial components of the gauge field $a_T^i$ to each other, so the equation (2.8) governs a single quadratically dispersing excitation.

In fact, it's possible to see the relation to a scalar theory directly, by making the non-local change of variables (setting $g = 1$)

$$\begin{aligned} e_i &= -\epsilon_{ij} \frac{\partial_j \partial_t \phi}{\nabla^2} - \frac{k}{2\pi} \partial_i \phi; \\ b &= \nabla^2 \phi \end{aligned} \qquad (2.9)$$

resulting in an action of the form (1.7) with the replacement $\kappa^2 \to \kappa^2(1 + \frac{\kappa^2 k^2 g^4}{4\pi}) = \tilde{\kappa}^2$. In contrast to the Lorentz-invariant case, where perturbing the Maxwell theory by a Chern-Simons term results in a massive theory [12], here the Chern-Simons term does *not* induce a mass for the photon. [3]

The breaking of parity and time-reversal invariance by the Chern-Simons coupling must be reflected in some property of the excitations of the Lifshitz-Chern-Simons theory. The mapping (2.9) seems to contradict this expectation. The same apparent contradiction occurs in the case of Maxwell-Chern-Simons theory, where a similar transformation maps the gauge theory to that of a free massive boson. However, the excitations described by the boson have a unit of spin which breaks parity [12]. This must occur in the Lifshitz-Chern-Simons theory as well.

A well-defined notion of spin in 2+1 dimensions requires the full structure of the non-abelian Lorentz group $SO(2,1)$. In the relativistic Maxwell-Chern-Simons theory, the spin operator, which naively is just the $SO(2)$ rotation generator, must be augmented by a new term (only present for non-zero Chern-Simons level). This additional term is found by the requirement that the full Lorentz algebra be conventional; commutators of boost generators would have been singular without such a term. The spin operator $M$ in momentum space takes the form

$$M = \int d^2p \; a^\dagger(p)(-i\frac{\partial}{\partial \theta})a(p) + \frac{k}{|k|}\int d^2p \; a^\dagger(p)a(p) \qquad (2.10)$$

where $\theta = \tan^{-1}(p_2/p_1)$, $a(p)$ ($a^\dagger(p)$) are the lowering (raising) operators in the mode expansion for the excitations, and the limit for zero Chern-Simons level is taken so that $k/|k|$ vanishes. It is the second term in (2.10) that is added to the usual rotation generator $\int x_i e_i b$ so that the commutators of boost generators are non-singular. This second term is obtained by a phase rotation of the raising and lowering operators [12].

Because the Lifshitz-Chern-Simons theory is non-relativistic, the non-abelian $SO(2,1)$ structure is not present. So there is no a priori reason to modify the rotation generator. But there does exist an RG flow initiated by the $e^2$ operator that drives the theory into a massive phase described by the Maxwell-Chern-Simons theory. Therefore, we define spin in the Lifshitz-Chern-Simons theory by its definition in the relativistic Maxwell-Chern-Simons theory, i.e., by the endpoint of this possible RG trajectory. The precise form of the spin operator is again given by (2.10), and it is clear that the second term measures one unit of spin for any gapless excitation.

---

[3] If we were to consider a compact U(1) gauge theory of this form, it is conceivable that such a mass could be induced non-perturbatively. Since our focus is on a non-compact version of the theory, we do not discuss monopole events here. In fact, the shift of $\kappa^2$ by the Chern-Simons coefficient lowers the dimension of monopole vertex operators [8,6]. See [14] for an interesting recent study on the effects of instantons in a related compact $U(1)$ $z=2$ gauge theory that is the 2+1 dimensional version of the abelian theory in [15].

*2.2. Ground State Degeneracy at the Critical Point*

When the Maxwell-Chern-Simons theory is quantized on $R \times \Sigma$, where $\Sigma$ is a genus $g$ Riemann surface, the presence of the Chern-Simons term results in degenerate ground states [16]. (There is no loss of generality in considering three-manifolds with the $R \times \Sigma$ product structure, since we may construct more general three-manifolds by gluing together these basic units with appropriate twists.) The degeneracy arises by quantizing the moduli space of Wilson lines of the gauge field around the one-cycles of the Riemann surface. For a genus $g$ Riemann surface, there are $2g$ one-cycles and the ground state degeneracy is $k^g$ where $k$ is the Chern-Simons coupling or level. The $k^g$ degeneracy provides an alternate explanation for the quantization of the abelian Chern-Simons level.

This result is obtained as follows [16]. If we are interested in the ground state, then we may restrict to zero momentum modes.[4] Restricting to zero momentum turns the quantum field theory problem into one of quantum mechanics. For the Maxwell-Chern-Simons theory (which applies to the quantum Hall phase on one side of the Lifshitz-Chern-Simons critical point), the quantum mechanics problem is that of $g$ particles moving on a two-dimensional plane in a magnetic field of strength $k$. To see this, write $a_i = a_i(t)$ where we are working in $a_t = 0$ gauge, and let us take $\Sigma = T^2$, the torus, for which $g = 1$, with unit length cycles. A momentum-independent gauge field clearly solves Gauss' law assuming there are no local sources. Plugging the ansatz for $a_i$ into the Maxwell-Chern-Simons action we find

$$S = \int dt \Big[ \frac{1}{2r\Lambda} (\partial_t a_i)^2 + k(a_1 \partial_t a_2 - a_2 \partial_t a_1) \Big]$$

where $\Lambda$ is the UV cutoff and $r$ is the dimensionless Maxwell coupling constant. This Landau level problem has a series of degenerate energy levels of degeneracy $k$ for each particle with a level separation of $kr\Lambda$. Taking the cutoff $\Lambda$ to infinity decouples the excited states. This excitation gap coincides with the analysis of [12].

---

[4] This is true assuming that the expansion about zero momentum is stable, i.e. that the vacuum does not spontaneously break translation invariance, which is true for the Maxwell-Chern-Simons theory. More generally, if this is not the case, then the theory must be expanded about the vacuum in which the unstable momentum modes have condensed.

If the gauge field is coupled to massive matter fields, then the degeneracy will not be exact in a finite-sized system. If $L$ is the length of the shortest loop around the torus, then the degeneracy will be split by $\Delta E \sim e^{-cmL}$ for some constant $c$ if $m$ is the mass of the lightest matter field [16]. As the mass becomes large and/or the length $L \to \infty$, the splitting becomes exact.

We now turn to the Lifshitz-Chern-Simons theory. Since the theory is gapless, the ground state degeneracy must be defined with care. The most natural way to do this is to consider the system on a torus of length $L$. Then, the excitations of the massless gauge field will have a gap $\propto 1/L^2$ and any ground state degeneracy can be cleanly defined and computed. If we were to couple the gauge field to massive matter fields, then the degeneracy would not be exact, but would be split by $\Delta E \sim e^{-\alpha L}$ for some $\alpha$. For large $L$, this splitting will be much smaller than $1/L^2$, so we can still distinguish degenerate ground states from excitations of the gauge field.

Using a finite size to separate low-energy excitations from degenerate ground states is already necessary in the usual quantum Lifshitz theory without Chern-Simons term:

$$S = \int dt d^2x \Big( e_i \partial_t a_i + a_t \partial_i e_i - \frac{1}{2}(\partial_i e_j)^2 - \frac{1}{2}b^2 \Big). \tag{2.11}$$

At the critical point described by this action, any constant electric field is allowed. Thus, there is a moduli space of vacua parametrized by constant $e_i$. (In the quantum dimer model, this corresponds to the fact that any integer winding number can occur at the Rokhsar-Kivelson point [5].) In addition, there are gapless excitations with $\omega \propto k^2$ dispersion. In a finite-size system, there is a gap $\propto 1/L^2$, so these excitations can be cleanly separated from the degenerate ground states.

We note, however, that if an $(e_i^2)^2$ term is included in the quantum Lifshitz action (2.11), then this degeneracy is lifted. In the quantum dimer model, such a term is forbidden, but it is allowed in the present context. This operator will be important in what follows.

We analyze the Abelian Lifshitz-Chern-Simons theory on the torus in a similar manner to the above analysis of Maxwell-Chern-Simons theory:

$$S = \int dt d^2x \Big( e_i \partial_t a_i + a_t \partial_i e_i - \frac{1}{2}(\partial_i e_j)^2 - \frac{\lambda}{4}(e_i^2)^2 - \frac{1}{2}b^2 + \frac{k}{4\pi}\epsilon_{\mu\nu\rho}a_\mu \partial_\nu a_\rho \Big). \tag{2.12}$$

For simplicity, we have set the coupling constants $g, \kappa^2$ introduced earlier to unity. We have also retained the marginally-irrelevant operator $(e_i^2)^2$ because we will shortly need it to lift the degeneracy between different constant values of $e_i$. Focusing again on the zero-momentum modes, we have:

$$S = \int dt \Big( e_i \partial_t a_i - \frac{\lambda}{4L^2}(e_i^2)^2 + \frac{k}{4\pi}(a_2 \partial_t a_1 - a_1 \partial_t a_2) \Big). \tag{2.13}$$

Here, we have used the Fourier transform convention $e_i(x) = \frac{1}{L} \sum_{\mathbf{p}} e_i(p) e^{-i\mathbf{p} \cdot \mathbf{x}}$, as a result of which the second term in (2.13) has explicit $L$ dependence. Regardless of the particular Fourier transform convention which we use, the ratio of the second term to the first and third terms will be $L$-dependent for dimensional reasons because it does not have a time derivative. Using the canonical momenta (2.3), we find the Hamiltonian for the zero-momentum modes to be:

$$H = \frac{\lambda}{4L^2} \left[ \left( \Pi_i - \frac{k}{4\pi} \epsilon_{ij} a_j \right)^2 \right]^2. \tag{2.14}$$

This is just the square of the usual Hamiltonian for a particle in a magnetic field. Thus, the spectrum is simply a series of Landau levels but, rather than being evenly-spaced, $E_n \propto (n + \frac{1}{2})$, they are quadratically-spaced, $E_n \propto (n + \frac{1}{2})^2/L^2$. The key fact for us is that the ground state has degeneracy $k$ due to the degeneracy of the lowest Landau level. If we had dropped the $(e_i^2)^2$ term, the splitting between Landau levels would have collapsed and, in addition to the degeneracy within each Landau level, we would have had an additional (infinite) degeneracy between all of the Landau levels. However, for $\lambda/L^2$ finite, these higher Landau levels are separated from the degenerate ground states in the lowest Landau level.

Thus, we see that we have a $k$-fold degenerate ground state for Lifshitz-Chern-Simons theory on the torus and, by a straightforward extension of these arguments, a $k^g$-fold degeneracy on higher-genus surfaces. So long as there is a finite $(e_i^2)^2$ term, there is no further degeneracy beyond this. There is a gap $\propto 1/L^2$ to the lowest excited state above this degenerate ground state subspace. If matter fields with a non-zero mass gap are included, we expect an exponentially-small splitting of the degenerate ground states, as in the Maxwell-Chern-Simons case. This splitting is much smaller than the finite-size gap $\propto 1/L^2$.

## 2.3. Edge Modes at the Critical Point

Under a gauge transformation $a_\mu \to a_\mu + \partial_\mu \alpha$, $e_i \to e_i$, the abelian Lifshitz-Chern-Simons Lagrangian (2.12) is invariant up to a boundary term

$$\delta L = \frac{k}{4\pi} \int dt d^2x \; \partial_\mu(\epsilon_{\mu\nu\rho}\alpha\partial_\nu a_\rho) \tag{2.15}$$

which vanishes as long as the gauge curvature $f_{\mu\nu} = \partial_\mu a_\nu - \partial_\nu a_\mu$ and gauge variation parameter $\alpha$ asymptote to zero sufficiently quickly at infinity. When the three-manifold has a boundary, this boundary term is no longer necessarily zero. In order for the theory to be well-defined, it is necessary to add chiral edge modes that propagate along the 1+1 dimensional boundary (assuming a spatial boundary) that soak up any would-be gauge anomaly [17,18,19,20].

When the theory describing the bulk is simply Maxwell-Chern-Simons theory, it is clear that the chiral edge modes persist because at low energies the gapped bulk modes decouple and cannot cause any back scattering. In the Lifshitz-Chern-Simons theory, the bulk theory is gapless, so it's natural to wonder if the edge modes are still stable to perturbations.

Two simple arguments indicate that the edge modes remain protected. The first is a nonperturbative argument. The existence of the edge modes is tied to the gauge anomaly in (2.15). If interactions between the gapless bulk and the chiral edge modes somehow lifted the edge degrees of freedom, the edge excitations would no longer soak up the would-be gauge anomaly, and the theory would be ill-defined.

The second reason is more constructive. The Lagrangian that describes the chiral bosonic edge degree of freedom $\phi$ is given by

$$S = \frac{k}{4\pi} \int dtdx \; \partial_x\phi(i\partial_t - \partial_x)\phi. \tag{2.16}$$

The global $U(1)$ Kac-Moody symmetry of (2.16) is identified with the restriction of the bulk $U(1)$ gauge symmetry. Gauge-invariant minimal couplings between the edge mode and the restriction of the bulk gauge field have been suppressed in (2.16). The vertex operator $e^{i\phi}$ creates an edge excitation of charge $1/k$. Because the quadratically dispersing bulk mode is neutral, we cannot form a gauge-invariant relevant operator coupling the bulk mode to the chiral edge mode. Even disorder cannot change this conclusion [21]. In fact, the free chiral conformal field theory (2.16) admits no relevant perturbations at all.

## 3. Anisotropic Phase

*3.1. Clean Systems*

In the past few sections, we have focused on the critical point at which the coefficient of the $e_i^2$ term is zero. Suppose we move away from the critical point. Then the action takes the form:

$$S = \int dt\, d^2x \left[ \frac{1}{g^2}\left(e_i\partial_t a_i + a_t \partial_i e_i - \frac{r}{2}e_i^2 - \frac{\kappa^2}{2}(\partial_i e_j)^2 - \frac{\lambda}{4}(e_i^2)^2 - \frac{1}{2}b^2\right) + \frac{k}{4\pi}\epsilon_{\mu\nu\rho}a_\mu \partial_\nu a_\rho \right]. \quad (3.1)$$

As noted previously, for $r > 0$, we can integrate out $e_i$, thereby obtaining Maxwell-Chern-Simons theory. The Maxwell term is irrelevant, so this theory is dominated in the IR by the Chern-Simons term. This is just the effective theory of a quantum Hall state, as we will see in the next section, where we compute response functions.

For $r < 0$, $e_i$ develops an expectation value, thereby breaking SO(2) rotational symmetry. To see this, note that for $r < 0$, the Hamiltonian

$$g^2 H[e,a] = -\frac{|r|}{2}e_i^2 + \frac{\kappa^2}{2}(\partial_i e_j)^2 + \frac{\lambda}{4}(e_i^2)^2 + b^2 \quad (3.2)$$

is minimized by $e_i^2 = \frac{|r|}{\lambda}$, $a_i = 0$. Without loss of generality, we can choose the SO(2) symmetry-breaking vacuum to be given by $e_i = (\sqrt{|r|/\lambda}, 0)$. We expand about this point by writing $e_i = (\sqrt{|r|/\lambda} + \tilde{e}_x, e_y)$ and substitute this into (3.2), thereby obtaining

$$H[e,a] = |r|\tilde{e}_x^2 + \frac{\kappa^2}{2}(\partial_i e_y)^2 + b^2 + \mathcal{O}(e^3), \quad (3.3)$$

where we have dropped an unimportant constant and truncated the Hamiltonian to the lowest dimension operators. Since the symmetry-breaking electric field is constant, we did not need to worry about derivative terms in the electric field.

The corresponding Lagrangian is then:

$$S = \int dt\, d^2x \left[ \frac{1}{g^2}\left(\sqrt{|r|/\lambda}\,\partial_t a_x + \tilde{e}_x \partial_t a_x - |r|\tilde{e}_x^2 - (\partial_y a_x - \partial_x a_y)^2 + a_t \partial_i e_i \right.\right.$$
$$\left.\left. + e_y \partial_t a_y - \frac{\kappa^2}{2}(\partial_i e_y)^2\right) + \frac{k}{4\pi}\epsilon_{\mu\nu\lambda}a_\mu \partial_\nu a_\lambda \right]. \quad (3.4)$$

The linear term proportional to $\partial_t a_i$ can be removed by the shift $a_y \to a_y - \frac{4\pi}{k}\sqrt{\frac{|r|}{\lambda}}$. This theory is gapless since it must contain the Goldstone boson of broken rotational symmetry; it is expected to have $z = 1$ since the order parameter is not conserved [22]. We can check

that this theory does, indeed have, $z = 1$ by noting that the terms in the action (3.4) which dominate at long wavelengths are invariant under the scaling $t \to \lambda t$, $\mathbf{x} \to \lambda \mathbf{x}$, provided we scale:

$$\begin{aligned} e_x &\to \lambda^{-3/2}\, e_x\,, & e_y &\to \lambda^{-1/2}\, e_y \\ a_x &\to \lambda^{-1/2}\, a_x\,, & a_y &\to \lambda^{-3/2}\, a_y\,, & a_t &\to \lambda^{-3/2}\, a_t \end{aligned} \quad (3.5)$$

(with the non-invariant terms giving irrelevant corrections to the leading long-wavelength behavior). This can also be seen directly from the propagators. For instance,

$$\begin{aligned} \langle e_x(-i\omega_n, -\mathbf{p})\, e_x(i\omega_n, \mathbf{p}) \rangle &= \frac{g^2(p_y^2 + \kappa^2 g^4 \tilde{k}^2 p^2)}{\omega_n^2 + g^2 r p_y^2 + \kappa^2 p^2 (g^6 \tilde{k}^2 r + p_x^2)} \\ \langle e_y(-i\omega_n, -\mathbf{p})\, e_y(i\omega_n, \mathbf{p}) \rangle &= \frac{g^2(g^6 \tilde{k}^2 r + p_x^2)}{\omega_n^2 + g^2 r p_y^2 + \kappa^2 p^2 (g^6 \tilde{k}^2 r + p_x^2)} \end{aligned} \quad (3.6)$$

Thus, there is a gapless Goldstone mode with

$$\omega^2 = (\kappa^2 g^6 \tilde{k}^2 r) p_x^2 + (g^2 r + \kappa^2 g^6 \tilde{k}^2 r) p_y^2 + \kappa^2 p^2 p_x{}^2 \quad (3.7)$$

The third term is subleading at low energies. The first two terms describe a $z = 1$ mode with anisotropic velocities:

$$v_x^2 = \kappa^2 g^6 \tilde{k}^2 r, \quad v_y^2 = g^2 r + \kappa^2 g^6 \tilde{k}^2 r \,. \quad (3.8)$$

In the next section, we will see how this Goldstone mode contributes to response functions. Note that, for Chern-Simons level $k = 0$, $\omega^2 = g^2 r p_y^2 + \kappa^2 p_x{}^4$, so $p_x$ and $p_y$ scale differently; for non-zero $k$, the velocities are different, but $p_x$ and $p_y$ scale the same way.

It is important to note, however, that rotational symmetry will not be an exact symmetry of most systems. Disorder will explicitly violate rotational symmetry, as will the underlying crystalline lattice. Since the main application which we have in mind is the quantum Hall effect in semiconductor heterostructures or quantum wells, we expect the influence of the lattice to be very weak. For this reason, we neglected terms which respect only a discrete rotational symmetry (and which are important in the quantum dimer model [4,6]). Since such terms, if present, would have very small bare coefficients, and they are at best marginally-relevant (which is the scaling found at $k = 0$ [4,6]), they would only grow logarithmically. Though they might alter the nature of the transition, possibly even driving it first-order, these effects would only be apparent very close to the transition point. On the other hand, these SO(2)-violating terms are strongly-relevant in the symmetry-broken phase and make an immediate impact there. Furthermore, lattice effects are simpler to study than disorder (to which we turn in the next subsection) and they illustrate the important point that the Goldstone boson found above is not generic.

Therefore, we consider the Hamiltonian

$$g^2 H[e,a] = -\frac{|r|}{2}(e_x^2 + (1-\alpha^2)e_y^2) + \frac{\kappa^2}{2}(\partial_i e_j)^2 + \frac{\lambda}{4}(e_i^2)^2 + b^2 ,  \qquad (3.9)$$

where $\alpha^2$ is an anisotropy parameter which vanishes at the isotropic point and is greater or less than 0 if the stiffness in the $y$-direction is, respectively, greater or less than that in the $x$-direction. (In general, the other terms will also be anisotropic, but the leading quadratic term is the most important term, so for the sake of simplicity, we will consider the case in which only this term is anisotropic.) Without loss of generality, let us suppose that $\alpha^2 > 0$. While this action is no longer O(2) invariant, it is still $D_2$-invariant, and this symmetry is spontaneously broken to $Z_2$ when $e_x$ develops an expectation value. Then we expand the fields about the minimum of the Hamiltonian $e_i = (\sqrt{|r|/\lambda} + \tilde{e}_x, e_y)$, which leads to the Hamiltonian

$$H[e,a] = |r|\tilde{e}_x^2 + |r|\alpha^2 e_y^2 + b^2 + \mathcal{O}(e^3) . \qquad (3.10)$$

Here, we have kept only the most relevant terms. The fields $\tilde{e}_x$ and $e_y$ can now be integrated out of the action to give:

$$S = \int dt\, d^2x \left[ \frac{1}{g^2} \left( \sqrt{|r|/\lambda}(\partial_t a_x - \partial_x a_t) + \frac{1}{|r|}(\partial_t a_x - \partial_x a_t)^2 + \frac{1}{|r|\alpha^2}(\partial_t a_y - \partial_y a_t)^2 \right.\right.$$

$$\left.\left. - (\partial_y a_x - \partial_x a_y)^2 \right) + \frac{k}{4\pi}\epsilon_{\mu\nu\lambda} a_\mu \partial_\nu a_\lambda \right].$$

(3.11)

Again, the first term can be eliminated by shifting $a_y \to a_y - \frac{4\pi}{k}\sqrt{|r|/\lambda}$. The resulting action is simply anisotropic Maxwell-Chern-Simons theory and has a gap. Consequently, this phase will have protected edge modes and a $k$-fold ground state degeneracy precisely as in the $r > 0$ phase.

Thus, in the absence of a continuous rotational symmetry, the (discrete) symmetry-breaking phase is also a quantum Hall state with Hall conductance $1/k$, precisely as it is on the other side of the transition. However, the irrelevant Maxwell terms are anisotropic. Thus, we expect finite-frequency and, in the presence of matter fields, finite-temperature transport to be anisotropic. When the Lagrangian is SO(2) rotation-symmetric, there is, in addition, a Goldstone boson. In the next section, where we compute the response functions in both phases and the critical point, we will see how the Goldstone mode affects transport properties. First, however, we consider the effects of disorder.

## 3.2. Dirty Systems

In any real system, there will be impurities, and they will have the effect of changing the couplings in the action into random functions of the spatial position (constrained only by the symmetries respected by the impurities). In this section, we will not try to systematically determine the effect of disorder by allowing all possible couplings to be random, which would be important in order to determine the true ground state of the system for $r < 0$. Instead, we will examine one particular random coupling in order to see some of the generic qualitative features of disorder. Let us suppose that the coefficients of $e_x^2$ and $e_y^2$ are random:

$$S = \int dt\, d^2x \left[ \frac{1}{g^2}\left(e_i \partial_t a_i + a_t \partial_i e_i - \frac{1}{2}re_i^2 - \frac{1}{2}r_x(\mathbf{x})e_x^2 - \frac{1}{2}r_y(\mathbf{x})e_y^2 + \ldots \right] . \quad (3.12)$$

The ... are the other terms which are present (3.1). Although they, too, can be random, we assume, for illustrative purposes, that all other couplings in the theory are constant. Here, we have separated the mean value $r$ of the coefficients of $e_x^2$ and $e_y^2$ from the random parts, $r_x(\mathbf{x})$ and $r_y(\mathbf{x})$, which we assume to be Gaussian white-noise correlated with zero mean and variance $W_{x,y}$:

$$\overline{r_i(\mathbf{x})} = 0, \quad \overline{r_i(\mathbf{x}) r_j(\mathbf{x}')} = W_i \delta_{ij}\, \delta(\mathbf{x} - \mathbf{x}'). \quad (3.13)$$

Thus, the system is isotropic on average, but in any given realization of the disorder, the system is anisotropic. We can compute disorder-averaged correlation functions by replicating the theory (i.e. taking $n$ copies of the theory), integrating out $r_x(\mathbf{x})$ and $r_y(\mathbf{x})$, and taking the replica limit $n \to 0$. (The replica trick ensures that we are computing correlation functions from derivatives of the disorder-averaged free energy rather than the disorder-averaged partition function.) This leads to the replicated action

$$\begin{aligned}
S = &\int dt\, d^2x \left[ \frac{1}{g^2}\left(e_i^A \partial_t a_i^A + a_t^A \partial_i e_i^A - \frac{1}{2}r(e_i^A)^2 + \ldots \right] \\
&+ \int dt\, dt'\, d^2x \left[ -\frac{1}{4g^4} W_x (e_x^A(x,t))^2 (e_x^B(x,t'))^2 - \frac{1}{4g^4} W_y (e_y^A(x,t))^2 (e_y^B(x,t'))^2 \right].
\end{aligned}$$
$$(3.14)$$

where $A = 1, 2, \ldots, n$ is the replica index.

We now expand about a minimum of the potential energy of the clean part of the action, $e_i = (\sqrt{|r|/\lambda}, 0)$ by writing $e_i = (\sqrt{|r|/\lambda} + \tilde{e}_x, e_y)$ and substitute this into (3.14), to obtain

$$S = \int dt\, d^2x \left[\frac{1}{g^2}\left(\sqrt{|r|/\lambda}\, \partial_t a_x^A + \tilde{e}_x^A \partial_t a_x^A - |r|(\tilde{e}_x^A)^2 - (\partial_y a_x^A - \partial_x a_y^A)^2 + a_t^A \partial_i e_i^A \right.\right.$$
$$\left.\left. + e_y^A \partial_t a_y^A - \frac{\kappa^2}{2}(\partial_i e_y^A)^2\right) + \frac{k}{4\pi}\epsilon_{\mu\nu\lambda} a_\mu^A \partial_\nu a_\lambda^A\right]$$
$$+ \int dt\, dt'\, d^2x \left[-\frac{1}{4g^4} W_x (e_x^A(x,t))^2 (e_x^B(x,t'))^2 - \frac{1}{4g^4} W_y (e_y^A(x,t))^2 (e_y^B(x,t'))^2\right]. \tag{3.15}$$

Here, we have allowed $W_x$ and $W_y$ to be different since they flow differently even if they are the same microscopically: using the scaling (3.5), we see that

$$W_x \to \lambda^{-2} W_x, \qquad W_y \to \lambda^2 W_y. \tag{3.16}$$

Thus, $W_x$ is strongly irrelevant while $W_y$ is strongly relevant.

The relevance of $W_y$ is a symptom of the fact that, in the presence of disorder, no ordered state is possible. This is because, according to the Imry-Ma argument [23], a reversed domain of linear size $L$ will cost energy $\sim L^{d-2}$, assuming that the system is spatially-isotropic on average. On the other hand, it can lower its energy by $L^{d/2}$ by taking advantage of local fluctuations of the disorder. Thus, order is impossible in $d < 4$. (However, if the Hamiltonian only has a discrete rotational symmetry on average, then a reversed domain will cost energy $\sim L^{d-1}$, and an ordered state at $r < 0$ is possible in two spatial dimensions.)

Let us focus on systems which are spatially-isotropic on average. In such systems, there is a length scale, the Larkin length $\xi_L$, at which the energy cost for a reversed droplet $\sim \xi_L^{d-2}$ is equal to the energy gain $\sim \xi_L^{d/2}$. At shorter length scales, the system is aligned in the direction chosen by the disorder. At longer scales, the system breaks up into different domains of size $\xi_L$ which are aligned in different directions.

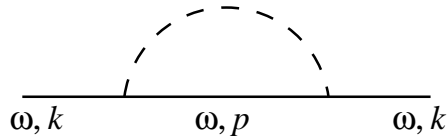

**Fig. 1.** Lowest order diagram contributing to the $e_y$ self-energy.

There is a corresponding energy scale, the pinning energy $E_p$. At energies higher than $E_p$, we can compute perturbatively in $W_y$. The lowest order diagram contributing to the $e_y$ self-energy is in Figure 1. This diagram arises at second order in the cubic interaction between $e_y$ and $r_y$ in (3.12). The dotted line represents the disorder average of $r_y$ while the solid line is the $e_y$ propagator. Its value is

$$\Sigma(\omega, \mathbf{k}) = W_y \int \frac{d^2p}{(2\pi)^2} \frac{g^2(g^6\tilde{k}^2 r + p_x^2)}{\omega^2 - g^2 r p_y^2 - \kappa^2 p^2(g^6\tilde{k}^2 r + p_x^2) + i\delta}. \tag{3.17}$$

Note that there is no internal frequency integral because the disorder is time-independent and interacts elastically. $\mathbf{k}$ dependence would be possible had we chosen a different averaging scheme than (3.13). The imaginary part of the self-energy has a constant $\omega$-independent piece which is:

$$\text{Im } \Sigma(\omega, \mathbf{k}) = \frac{W_y g^8 \tilde{k}^2 r}{2 v_x v_y} + \ldots \tag{3.18}$$

where $v_x, v_y$ are the velocities (3.8). We focus on this piece of $\Sigma(\omega, \mathbf{k})$ since it is qualitatively important for the transport properties which we compute in the next section. There is also an $\omega$-dependent term (which will not be important for us), and a real part which diverges logarithmically at small $\omega$ (as expected by applying the Kramers-Kronig relation to (3.18)). This log divergence is also expected because, as we have seen, the coupling to disorder is a dimension-2 relevant operator, which prevents the existence of a symmetry-breaking ordered state by the arguments given above.

## 4. Response Functions at and near the Critical Point

In this section, we determine how the system responds to an external gauge field by calculating various response functions at the critical point and in the phases on either side of it. We will calculate both static and dynamic response functions. Static and dynamic response is distinguished by the order in which the zero frequency and zero momentum limits are taken. For static response the frequency is taken to zero first and followed by the zero momentum limit, while this ordering is switched for dynamic response. Such limits need not commute.

The general outline of the response calculation is the following. First, we minimally couple the external gauge field $A_\mu$ to the current in (1.2),[5]

$$\delta L = \int dt d^2x \; J^\mu A_\mu. \tag{4.1}$$

We take both the internal $a_\mu$ and electromagnetic $A_\mu$ gauge fields to be in the transverse gauge, $\partial_i a_i = 0$. This constraint fixes time-independent gauge transformations. Time-dependent gauge transformations are fixed by either setting $a_t$ to zero or a frequency-independent function, depending upon which choice is most convenient. Because the action is quadratic, we can integrate out the $a_\mu$ gauge field. In momentum space, the action for the external gauge field then takes the form

$$S_{\text{eff}} = \frac{1}{2} \sum_n \int d^2p \; A_\mu(-i\omega_n, -\mathbf{p}) K_{\mu\nu}(i\omega_n, \mathbf{p}) A_\nu(i\omega_n, \mathbf{p}) \tag{4.2}$$

for some matrix $K_{\mu\nu}(i\omega_n, \mathbf{p})$. $\text{Re}(K_{\mu\nu})$ must be symmetric while $\text{Im}(K_{\mu\nu})$ is antisymmetric in the indices $\mu\nu$. Further, gauge invariance of (4.2) implies that $p^\mu K_{\mu\nu}(i\omega_n, \mathbf{p})$ is zero. These requirements are satisfied by the $K_{\mu\nu}$ found in different regimes below. The conductivity is then obtained from $K_{\mu\nu}$ according to:

$$\sigma_{jk}(\omega) = \frac{1}{i\omega + \delta} K_{jk}(\omega + i\delta, \mathbf{p} = 0), \tag{4.3}$$

where $\omega, p$ is the frequency and momentum of the external field. The analytic continuation $i\omega_n \to \omega + i\delta$ prescription ensures that the right-hand-side is given by the retarded Green function.

Since the current operator is given by $J_\mu = \frac{1}{2\pi} \epsilon_{\mu\nu\rho} \partial_\nu a_\rho$, $K_{\mu\nu}(i\omega_n, p)$ is given by:

$$K_{\mu\nu}(i\omega_n, \mathbf{p}) = \left(\frac{1}{2\pi}\right)^2 \epsilon_{\mu\alpha\beta} \epsilon_{\nu\lambda\rho} p_\alpha p_\lambda \langle a_\beta(-i\omega_n, -\mathbf{p}) a_\rho(i\omega_n, \mathbf{p}) \rangle \tag{4.4}$$

### 4.1. Quantum Hall Phase at $r > 0$

We begin on familiar ground by computing the response functions on the isotropic quantum Hall side of the critical point. As noted earlier, for $r > 0$, $e_i$ can be integrated out, yielding Maxwell-Chern-Simons theory at low energy. Therefore, all response functions will be the same as in Maxwell-Chern-Simons theory, i.e. the system will be in a fractional quantum Hall phase. However, it is instructive to compute this in the full action (1.11). The expressions for the propagators are extremely cumbersome and not particularly enlightening, so we directly give the response functions:

$$K_{xx}(i\omega_n, \mathbf{p}) = \frac{1}{8\pi^2} \frac{g^2(g^4r^2p_y^2 + \kappa^2p^2(\omega_n^2 + \kappa^2p_y^2p^2) + g^2r(\omega_n^2 + 2\kappa^2p_y^2p^2))}{g^8\tilde{k}^2r^2 + \omega_n^2 + g^2rp^2 + 2\kappa^2g^2\tilde{k}^2rp^2 + \kappa^2p^4 + \kappa^4g^4\tilde{k}^2p^4}$$
$$K_{xy}(i\omega_n, \mathbf{p}) = \frac{1}{8\pi^2} \frac{g^2(g^2\tilde{k}\omega_n - p_xp_y)(g^2r + \kappa^2p^4)}{g^8\tilde{k}^2r^2 + \omega_n^2 + g^2rp^2 + 2\kappa^2g^2\tilde{k}^2rp^2 + \kappa^2p^4 + \kappa^4g^4\tilde{k}^2p^4}$$
(4.5)

---

[5] To avoid confusion, we should stress that as is conventional in the condensed matter literature, our $a_\mu$ is an "emergent" gauge field which is not to be confused with the electromagnetic field, which we represent by $A_\mu$.

Thus,

$$\sigma_{xx}(\omega) = \frac{i}{2\pi k} \cdot \frac{\omega/g^4\tilde{k}r}{1 - (\omega/g^4\tilde{k}r)^2}$$
$$\sigma_{xy}(\omega) = \frac{1}{2\pi k} \cdot \frac{1}{1 - (\omega/g^4\tilde{k}r)^2}$$
(4.6)

As expected, at low-frequencies, the real part of the longitudinal conductivity vanishes and the Hall conductivity is equal to $\frac{1}{2\pi k}$. There is a magnetoplasmon at frequency $\omega_{\rm mp} = g^4\tilde{k}r$. At higher frequencies, the conductivities fall off as $1/\omega$ and $1/\omega^2$ respectively. Note that the $\omega \to 0$ and $r \to 0$ limits don't commute.

The compressibility is obtained from the $\mathbf{p} \to 0$ limit of the static response function $K_{00}(0, \mathbf{p})$ and vanishes as $p^2$ in the $r > 0$ quantum Hall phase.

This computation could have been performed by integrating out $e_i$ and expanding the resulting action in powers of momentum. Only the Maxwell-Chern-Simons terms are necessary for the conductivities (4.6), but we would have needed to keep higher-order terms in order to obtain the full finite-frequency momentum expression (4.5).

*4.2. Critical Point at $r = 0$*

Next, we study the critical point described by Lifshitz-Chern-Simons theory. At the critical point, we have the propagators (1.9) computed earlier. Substituting the last three propagators in (1.9) into (4.4), we find

$$K_{xx}(i\omega_n, \mathbf{p}) = \frac{1}{8\pi^2} \frac{\kappa^2g^2p^2(\omega_n^2 + \kappa^2p_y^2p^2)}{\omega_n^2 + \tilde{\kappa}^2p^4}$$
$$K_{xy}(i\omega_n, \mathbf{p}) = \frac{1}{8\pi^2} \frac{\kappa^4g^2p^4(g^2\tilde{k}\omega_n - p_xp_y)}{\omega_n^2 + \tilde{\kappa}^2p^4}$$
(4.7)

Here, $\tilde{k} \equiv k/4\pi$, and $\tilde{\kappa}^2 = \kappa^2(1 + \kappa^2 \tilde{k}^2)$. Substituting these into (4.3), we find that since the $K_{xx}$, $K_{xy}$ are proportional to $p^2$ and $p^4$, respectively, both the real and imaginary parts of both conductivities vanish at any frequency:

$$\begin{aligned} \operatorname{Re} \sigma_{xx}(\omega) = \operatorname{Im} \sigma_{xx}(\omega) = 0 \\ \operatorname{Re} \sigma_{xy}(\omega) = \operatorname{Im} \sigma_{xy}(\omega) = 0 \end{aligned} \quad (4.8)$$

This is quite remarkable: the critical point is completely insulating – not merely at zero frequency, but at any frequency. By contrast, in the fractional quantum Hall phase on one side of the critical point, $\sigma_{xy} = 1/(2\pi k)$ and $\sigma_{xx} \propto i\omega$ at low-frequencies. Thus, as the gap collapses, the system becomes *more* insulating. The reason is that the critical theory doesn't couple to a spatially-homogenous electric field, regardless of the frequency (except at frequency precisely equal to zero).

The compressibility is obtained from the $\mathbf{p} \to 0$ limit of the static response function $K_{00}(0, \mathbf{p})$. From the propagators in propagators in (1.9), we see that

$$K_{00}(i\omega_n, \mathbf{p}) = \frac{1}{8\pi^2} \frac{\kappa^2 g^2 p^4}{\omega_n^2 + \tilde{\kappa}^2 p^4} \quad (4.9)$$

Thus, the compressibility at the critical point is given by

$$K = \frac{g^2}{8\pi^2(1 + \kappa^2 \tilde{k}^2)} \quad (4.10)$$

The critical point has larger (i.e. non-zero) compressibility than the quantum Hall phase (where it is zero).

We could have performed the computations in this subsection by integrating out $e_i$. The resulting action would have been non-local because it would have contained inverse powers of the momentum. However, the resulting singularities at zero momentum are not problematic because the external electromagnetic field does not couple to the zero momentum modes.

*4.3. Anisotropic Phase at $r < 0$*

If the system does not have SO(2) rotational invariance (e.g. only $D_2$ symmetry), then the low-energy effective theory in the anisotropic phase is anisotropic Maxwell-Chern-Simons theory (3.11). The response functions are then a slight modification of (4.5), and

the conductivities are:

$$\sigma_{xx}(\omega) = \alpha \frac{i}{2\pi k} \cdot \frac{\omega/g^4 \tilde{k} r \alpha}{1 - (\omega/g^4 \tilde{k} r \alpha)^2}$$
$$\sigma_{yy}(\omega) = \frac{1}{\alpha} \cdot \frac{i}{2\pi k} \cdot \frac{\omega/g^4 \tilde{k} r \alpha}{1 - (\omega/g^4 \tilde{k} r \alpha)^2} \quad (4.11)$$
$$\sigma_{xy}(\omega) = \frac{1}{2\pi k} \cdot \frac{1}{1 - (\omega/g^4 \tilde{k} r \alpha)^2}$$

At zero frequency, this is simply a quantum Hall state with Hall conductance $\sigma_{xy} = 1/2\pi k$. A similar calculation shows that the compressibility vanishes as $p^2$ as the momentum $\mathbf{p} \to 0$. The anisotropy shows up at finite frequency in the imaginary part of the longitudinal conductivity.

Now suppose that the system has the full SO(2) rotational symmetry so that there is a Goldstone boson in the anisotropic phase. Then, we have

$$K_{xx}(i\omega_n, \mathbf{p}) = \frac{1}{8\pi^2} \frac{\kappa^2 g^2 p^2 (\omega_n^2 + g^2 r p_y^2)}{\omega_n^2 + g^2 r p_y^2 + \kappa^2 p^2 (g^6 \tilde{k}^2 r + p_x^2)}$$
$$K_{yy}(i\omega_n, \mathbf{p}) = \frac{1}{8\pi^2} \frac{g^4 r (\omega_n^2 + \kappa^2 p_x^2 p^2)}{\omega_n^2 + g^2 r p_y^2 + \kappa^2 p^2 (g^6 \tilde{k}^2 r + p_x^2)} \quad (4.12)$$
$$K_{xy}(i\omega_n, \mathbf{p}) = \frac{1}{8\pi^2} \frac{\kappa^2 g^2 r (g^2 \tilde{k} \omega_n - p_x p_y) p^2}{\omega_n^2 + g^2 r p_y^2 + \kappa^2 p^2 (g^6 \tilde{k}^2 r + p_x^2)}$$

Thus, the system is a superconductor in one direction and an insulator in the other:

$$\sigma_{xx}(\omega) = 0$$
$$\sigma_{yy}(\omega) = \frac{g^4 r}{8\pi^2} \delta(\omega) + \frac{ig^4 r}{8\pi^2 \omega} \quad (4.13)$$
$$\sigma_{xy}(\omega) = 0$$

Meanwhile, the compressibility response function

$$K_{00}(0, \mathbf{p}) = \frac{1}{8\pi^2} \frac{g^2 (g^2 r p_y^2 + \kappa^2 p_x^2 p^2)}{\omega_n^2 + g^2 r p_y^2 + \kappa^2 p^2 (g^6 \tilde{k}^2 r + p_x^2)}$$
$$= \frac{1}{8\pi^2} \frac{g^2 p_y^2}{(1 + \kappa^2 g^4 \tilde{k}^2) p_y^2 + g^6 \tilde{k}^2 p_x^2} \quad (4.14)$$

depends on how the limit $\mathbf{p} \to 0$ is taken since the result is zero or non-zero for $p_x/p_y \to \infty$ or 0, respectively.

Thus we see that, if the system is SO(2)-symmetric and, therefore, has the corresponding Goldstone boson in the anisotropic phase, then it is maximally anisotropic: superconducting in one direction and insulating in the other. However, if the system only has the discrete rotational symmetry of an underlying rectangular lattice, then is is weakly anisotropic: a quantum Hall state with anisotropic finite-frequency dielectric constants (i.e. imaginary part of the conductivity).

Finally, we consider a dirty system at frequencies above the pinning energy. Then, computing the propagators with the self-energy (3.18), we find:

$$\sigma_{xx}(\omega) = \frac{1}{8\pi^2} \frac{ig^4 \omega}{g^8 \tilde{k}^2 r + i(\omega\tau)^2}$$
$$\sigma_{yy}(\omega) = \frac{1}{8\pi^2} \frac{g^4 r \omega \tau^2}{g^8 \tilde{k}^2 r + i(\omega\tau)^2} \quad (4.15)$$
$$\sigma_{xy}(\omega) = \frac{1}{2\pi k} \cdot \frac{1}{1 + i(\omega\tau)^2/g^8 \tilde{k}^2 r}$$

where

$$\frac{1}{\tau^2} = \frac{W_y g^8 \tilde{k}^2 r}{2 v_x v_y} \quad (4.16)$$

Note the compressibility now vanishes for finite $\tau$. At least at these frequencies, we once again have an anisotropic fractional quantum Hall state. However, unlike in a pure system with discrete rotational symmetry, there is a finite real part of the conductivity in both the $x$ and $y$ directions at finite frequency. Note that the frequency dependence differs along the two directions. The real part of the conductivity vanishes as $\omega^3$ along the $x$−direction while it vanishes linearly as $\omega$ along the $y$−direction.

## 5. Discussion

In this paper, we have studied a $z = 2$ abelian gauge theory with a Chern-Simons term. Previous investigations focused on either the $z = 2$ theory without Chern-Simons term, or on the Maxwell-Chern-Simons theory (which shares the same leading IR behavior as the pure Chern-Simons theory). In particular, the $z = 2$ abelian gauge theory, or abelian Lifshitz gauge theory, without Chern-Simons term has been previously studied in the context of the quantum dimer model, where it describes the Rokhsar-Kivelson point [24,4,6] between columnar and staggered phases of dimers. In that context, the lattice is extremely important, and the critical point is generally multi-critical. Such a theory also describes a transition between a uniform superconducting state and a ground state with non-zero supercurrent – a state in which the phase of the order parameter varies in the ground state, which is distinct from the FFLO state in which the amplitude of the order parameter is at non-zero wavevector. Similarly, the Maxwell-Chern-Simons theory has been extensively studied in the context of the fractional quantum Hall effect, where it serves as an effective theory for the gapped states at plateaus.

Remarkably, we find that both the abelian Lifshitz gauge theory and the Chern-Simons term are marginal with respect to the $z = 2$ scaling. Thus, the theory is gapless. It has the following general properties. The spectrum consists of a single quadratically-dispersing gapless spin-1 photon. When the theory is quantized on a non-trivial Riemann surface of genus $g$ and linear dimension $L$, there is a gap $\propto 1/L^2$ separating the excited states from the $k^g$-fold degenerate ground state, where $k$ is the Chern-Simons level. Further, there exist gapless edge modes that live along any boundaries of the system. These modes are robust to disorder and do not interact with the gapless $z = 2$ bulk modes.

This theory, which we have dubbed Lifshitz-Chern-Simons theory, describes a critical point to which we can tune the system by varying $r$, the coefficient of the square of the electric field $e_i^2$. The following phases lie on either side of this critical point. When $r$ is positive, the theory is described by Maxwell-Chern-Simons theory in the IR. For negative $r$, and in the absence of $SO(2)$ rotational symmetry-violating terms in the Hamiltonian, the ground state of the system spontaneously breaks $SO(2)$ and, therefore, there is a Goldstone boson. On the other hand, if the system only has a discrete rotational symmetry, the broken-symmetry state is gapped. Although, in either case, the broken symmetry state has a preferred direction, the low-energy theory of fluctuations about this broken-symmetry state is, to leading order, invariant under a reversal of this direction. Thus, this phase behaves as if it had nematic symmetry.

If the conserved current in this theory is interpreted as the electromagnetic current, then this theory describes a transition between a quantum Hall state (the $r > 0$ phase) and an anisotropic state (the $r < 0$ phase). Charge transport properties at the critical point and in the anisotropic phase are rather unique. For positive $r$, the transport matches that of Maxwell-Chern-Simons theory – the effective description of the Hall effect. When $r = 0$, the system insulates for all frequencies and has a finite non-zero compressibility. When $r < 0$ in a rotationally-invariant system, the system insulates along the ordered direction, while it has a delta function peak in its longitudinal conductivity, similar to a superconductor, along the orthogonal direction. If the Hamiltonian of the system has lower symmetry, then the system is in a quantum Hall state with the same Hall conductance as on the other side of the transition and anisotropic finite-frequency longitudinal conductivity.

The latter state is reminiscent of a recent experiment [25] in which a tilted field drives a transition from an isotropic quantum Hall state in the second Landau level (e.g. $\nu = 7/3$) to a quantum Hall state with anisotropic finite-temperature longitudinal transport. At $\nu = 5/2$, in contrast, a tilted field drives a transition [26,27] from a quantum Hall state to a nematic phase with unquantized $\sigma_{xy}$ and metallic $\sigma_{xx}$, $\sigma_{yy}$. (Such nematic phases have been the subject of considerable theoretical work; see, for example, [28,29] and references therein. It is unclear what relation, if any, those theories have to the anisotropic phase discussed here, which picks a preferred direction, but has nematic symmetry to leading order in the interaction strength.) Thus the recent findings come as a surprise. It is possible that these surprising transport coefficients are the result of inhomogeneity of the two-dimensional electron system. However, they have a natural explanation within Lifshitz-Chern-Simons theory without appealing to inhomogeneity, since in the presence of a crystalline lattice and disorder (which are certainly present in this system), the $r < 0$ phase is an anisotropic fractional quantum Hall state. This interpretation rests upon the assumption that the principal effect of the tilted field is the modification of the electronic wavefunctions in the direction perpendicular to the plane which, in the long-wavelength effective theory, has the result of driving the coupling $r$ negative. In fact, the in-plane magnetic field which drives the transition not only drives $r$ negative but also picks a preferred direction in the plane so that the transition is rounded. However, it may be a sharp crossover if the in-plane field is effectively a small symmetry-violating perturbation. Although our analysis, divorced from a microscopic model, cannot show that this is correct or even a reasonable assumption, there is, at least, a clear test. If Lifshitz-Chern-Simons theory is applicable to the transition seen in this experiment, then both the longitudinal and Hall conductivities should be strongly suppressed at the transition. In fact, in Lifshitz-Chern-Simons theory, they are strictly zero, but, as a result of the rounding mentioned above and, perhaps, the presence of disorder, there will be small non-zero conductivities at finite temperature at the critical point. In our simple model, we found anisotropic behavior in the zero temperature, finite frequency response at leading order. It would be interesting to analyze an interacting model to see if zero-frequency, finite-temperature transport shares some of the exotic properties found here.

**Acknowledgements**

We would like to thank Allan Adams, James Eisenstein, Eduardo Fradkin, Michael Levin, John McGreevy, Todadri Senthil, and Shivaji Sondhi for useful discussions. M.M. acknowledges the hospitality of the Kavli Institute for Theoretical Physics during the completion of this research. M.M. was supported in part by funds provided by the U.S. Department of Energy (D.O.E.) under cooperative research agreement DE-FG0205ER41360. S.K. was supported in part by the NSF under grant PHY-05-51164.